\documentclass[aps,floatfix,longbibliography,superscriptaddress,onecolumn,final,notitlepage]{revtex4-2}

\usepackage[utf8]{inputenc}
\usepackage[T1]{fontenc}
\usepackage{multirow}
\usepackage{subcaption}
\usepackage{tabularx}
\usepackage{graphicx,caption}
\usepackage{amsfonts,amsmath,amssymb}
\usepackage{comment}
\usepackage{lineno}
\usepackage{hyperref}
\usepackage[explicit]{titlesec}
\usepackage{cleveref} 

\newcommand{\be}{\begin{equation}}
\newcommand{\ee}{\end{equation}} 
\newcommand{\bea}{\begin{eqnarray}}
\newcommand{\eea}{\end{eqnarray}}

\newcommand{\bern}{\textrm{Bern}}

\newcommand{\pois}{\textrm{Pois}}

\renewcommand{\bf}[1]{\textbf{#1}} 

\newcommand{\f}[2]{\frac{#1}{#2}}

\newcommand{\ccup}[1]{\left\{#1\right\}}

\newcommand{\bup}[1]{\left(#1\right)}
\newcommand{\rup}[1]{\left[#1\right]}

\renewcommand{\ref}[1]{[\ref{#1}]}
 
\usepackage{algorithm,algorithmic} 
\usepackage[algoruled,algo2e,ruled]{algorithm2e}  
\usepackage{sidecap}
\usepackage{boldline}
\usepackage{setspace}
\usepackage{xcolor}

\usepackage[table]{}
\usepackage{array}
\usepackage{booktabs}
\usepackage{adjustbox}  
 \usepackage{natbib}
\usepackage{float}

\usepackage{comment}
\usepackage{lineno}
\usepackage{hyperref}
\usepackage[explicit]{titlesec}
\usepackage{cleveref}

\usepackage{xcolor}
\usepackage{tabu}
\usepackage{tikz}

\usepackage{subcaption}

\usepackage[normalem]{ulem} 
 
\newcommand{\ACD}{\mbox{{\small ACD}}}    
\newcommand{\polbooks}{\mbox{{\footnotesize POLBOOKS}}} 
\newcommand{\polblogs}{\mbox{{\footnotesize POLBLOGS}}} 
 
\usepackage{graphicx,caption} 
\usepackage{cleveref}
\crefname{equation}{Eq.}{Eqs.}
\crefname{section}{Sec.}{Secs.}
\crefname{figure}{Fig.}{Figs.}

\begin{document}




\title{Anomaly detection and  community detection  in networks}

\author{Hadiseh Safdari}
	\email{hadiseh.safdari@tuebingen.mpg.de} 
	\affiliation{Max Planck Institute for Intelligent Systems, Cyber Valley, Tuebingen 72076, Germany}

\author{Caterina De Bacco}
	\email{caterina.debacco@tuebingen.mpg.de} 
	\affiliation{Max Planck Institute for Intelligent Systems, Cyber Valley, Tuebingen 72076, Germany}

\begin{abstract} 
Anomaly detection is a relevant problem in the area of data analysis. In networked systems, where individual entities interact in pairs, anomalies are observed when pattern  of interactions deviates from patterns considered regular. Properly defining what regular patterns entail relies on developing expressive models for describing the observed  interactions. It is crucial to address anomaly detection in networks.    
Among the many well-known models for networks, latent variable models - a class of  probabilistic models - offer promising tools to capture the intrinsic features of the data.
In this work, we propose a probabilistic generative approach that incorporates domain knowledge, i.e., community membership, as a fundamental model for regular behavior, and thus flags potential anomalies deviating from this pattern.  In fact,  community membership serves  as the building block  of a null model to identify the regular interaction patterns.  The structural information is included  in the model through latent variables for community membership and anomaly parameter. The algorithm  aims at inferring these latent parameters and then output the labels identifying anomalies on  the network edges.
\end{abstract}

\maketitle 

\section{Introduction} 
Anomalies or outliers - deviations in the observed data so extreme as to arouse suspicion \cite{Hawkins1980}- form an unavoidable and problematic obstacle for data scientists. Over the past decades, anomalies have engendered a growing sense of concern in fields as varied as intrusion detection for network systems \cite{Hodge:2004aa,Iliofotou2007, Ding2012IntrusionA}, fraud detection in banking industry \cite{Ghosh1994CreditCF,Agarwal2005},  identifying fake users and events in communication networks, and medical condition monitoring \cite{Solberg2005DetectionOO}, to name a few.  Methods and techniques from various fields, such as statistics  \cite{Thottan2010}, data mining \cite{Caruso2007}, machine learning \cite{CATANIA20121822, Subba2016}, and  information theory \cite{AMARAL201780}, have been employed to address this problem.

Much of this growing body of work focuses on standard tabular datasets \cite{Pang2021, Jinghui_2017, HawkinsS_2002,chandola2009anomaly}. However, anomaly detection in network datasets, where many individuals interact in complex ways, has been lagging behind.  
In fact, in many complex systems the data is made of pairwise interactions between individuals, i.e., the only observed  information. For instance, in social networks, we know the nature of interactions  between the individuals, i.e., friendship, financial, but we may not have any metadata about the individuals. In this context, anomalies can merely be detected by considering the set of interactions, and measuring which  nodes or edges manifest an interaction pattern that is significantly different from that of their peers. In the case of online social networks, for example, advanced detection techniques that are independent of profile information are needed to detect fake profiles and malicious activities.\\ 

Our main objective in this work is to investigate the anomaly detection problem in networks. We consider the problem of observing a network that can have two possible, and different, mechanisms for edge formation; one involves the majority of the edges, whereas the other is an anomaly that we aim to detect. In other words, we have a regular pattern of interactions, and an anomaly. The latter belongs to a subset of interactions that deviates from the regular pattern. 

In many networked systems, in particular social networks, the interaction pattern is driven by community membership: individuals belong to groups and this determines how they interact \cite{Girvan2002,fortunato2010community}.  To properly detect anomalies, one should incorporate this insight to build a suitable null model that distinguishes between regular interactions and those instead relatable to malicious activities.  
Thus, we focus on networks that display community patterns as the regular mechanism.

Efforts have been made in this area and various models have been proposed for applying community detection approaches in anomaly detection.  For instance, Prado-Romero \emph{et al.} \cite{Prado2017} proposes  an adaptive method to detect anomalies using the most relevant attributes for each community. In general, most of the approaches focus on  attributed graphs  to predict anomalous behavior  \cite{Gao2010modeling, Mueller2013,Sultana2018,SAVAGE2014}. 
 
Probabilistic generative models are however a powerful approach to tackle community detection, as they allow to incorporate domain knowledge about how interactions arise into rigorous probabilistic models. However, they have been rarely used in the context of anomaly detection. Along these lines, \cite{bojchevski2018bayesian} propose a Bayesian model that combines network edges with additional information on nodes to identify anomalous nodes. Here, instead we do not assume any extra information being available  a priori  beyond the network structure.


\definecolor{darkgreen}{rgb}{0.0, 0.5, 0.0}
\definecolor{darkblue}{rgb}{0, 0.24, 0.64}
\definecolor{darkorange}{rgb}{1, 0.45, 0.0}
\definecolor{yellowgreen}{rgb}{0.6, 0.8, 0.4}

\begin{figure}[b]  \label{fig:vizexample}
  \captionsetup[subfigure]{justification=centering}
  \begin{subfigure}{.30\textwidth}
    \centering
    \begin{tikzpicture}[scale=1.3]
    
      \draw[draw=lightgray, rounded corners, thick] (-2.,1.) rectangle (2.,-1.5) {};
      \draw[draw=blue, fill=white, thick](-1.5,2) circle [radius=0.3] node (a) {$w$};
      \draw[draw=orange, fill=white, thick](0,2) circle [radius=0.3] node (b) {$\mu$};
      \draw[draw=orange, fill=white, thick](1.5,2) circle [radius=0.3] node (c) {$\pi$};
      \draw[draw=black, fill=white, thick](0,0) circle [radius=0.3] node (d) {$A_{ij}$};
      \draw[draw=blue, fill=white, thick](-1.5,0) circle [radius=0.3] node (e) {$u_i$};
      \draw[draw=blue, fill=white, thick](-1.45,-1) circle [radius=0.3] node (f) {$v_i$};

     \draw [draw=blue, ->,  >=latex, thick, shorten >= 4pt, shorten <= 7pt] (a) -- (d) {};
     \draw [draw=orange, ->,  >=latex, thick, shorten >= 6pt, shorten <= 6pt] (b) -- (c) {};
     \draw [draw=blue, ->,  >=latex, thick, shorten >= 6pt, shorten <= 6pt] (b) -- (a) {};
     \draw [draw=orange, ->,  >=latex, thick, shorten >= 4pt, shorten <= 7pt] (c) -- (d) {};
     \draw [draw=blue, ->,  >=latex, thick, shorten >= 2pt, shorten <= 5pt] (e) -- (d) {};
      \draw [draw=blue, ->,  >=latex, thick, shorten >= 0pt, shorten <= 3pt] (f) -- (d) {};

      \draw(-2.2,0) node [blue, rotate=90] {\footnotesize $Z_{ij} = 0$};
      \draw(2.2,0) node [orange, rotate=90] {\footnotesize $Z_{ij} \neq 0$};
      
    \end{tikzpicture}
    \caption[b]{Graphical model representation.}
    \label{fig:relations}
  \end{subfigure}
  \begin{subfigure}{.4\textwidth}
    \centering
    \begin{adjustbox}{width=0.6\linewidth}
    \begin{tikzpicture}[rotate=50, scale=1.5]

      \begin{scope}[every node/.style={color=black,fill=darkblue!50, circle,
                                      inner sep=6pt}]

        \node (a) at (-.5,4.8) {};
        \node (d) at (1.8,4.2) {};
        \node (c) at (-.3,2.8) {};
        \node (e) at (-2,1) {};
        \node (g) at (2,1.5) {};
      \end{scope}

      \begin{scope}[every node/.style={color=black,fill=darkblue!50, circle,
                                      inner sep=6pt}]

        \node (b) at (-2.5,3) {};
        \node (f) at (-0.8,0.8) {};
        \node (h) at (-.5,-1.5) {};
        \node (i) at (1.3,-.0) {};

      \end{scope}

      \begin{scope}[every edge/.style={draw=darkblue,very thick}]
        \path [-] (a) edge (c);
        \path [-] (c) edge (e);
        \path [-] (c) edge (d);
        \path [-] (d) edge (g);
        \path [-] (g) edge (c);
        \path [-] (a) edge (b);
      \end{scope}

      \begin{scope}[every edge/.style={draw=darkblue,very thick}]
        \path [-] (f) edge (h);
        \path [-] (i) edge (h);
        \path [-] (i) edge (f);
      \end{scope}
      \begin{scope}[every edge/.style={draw=darkblue, thick}]
        \path [-] (c) edge (b);
        \path [-] (f) edge (c);
      \end{scope}
      \begin{scope}[every edge/.style={draw=orange, thick}]
        \path [-] (d) edge (h); 
      \end{scope}

    \end{tikzpicture}
    \end{adjustbox}
    \vspace{0.2cm}
    \caption[b]{Network example.}
    \label{fig:example}
  \end{subfigure}
  \caption{Model visualization. (a) Graphical model: the entry of the adjacency matrix $A_{ij}$ is determined by the community-related latent variables $u, v, w$ (blue),  and by the anomaly  parameters $\pi$ (orange), depending on the values taken by the hyper-prior $\mu$. (b) Example of possible realization of the model: blue edges display interactions mainly based on the community, the orange edge exhibits an anomalous edges.\\}
\end{figure}
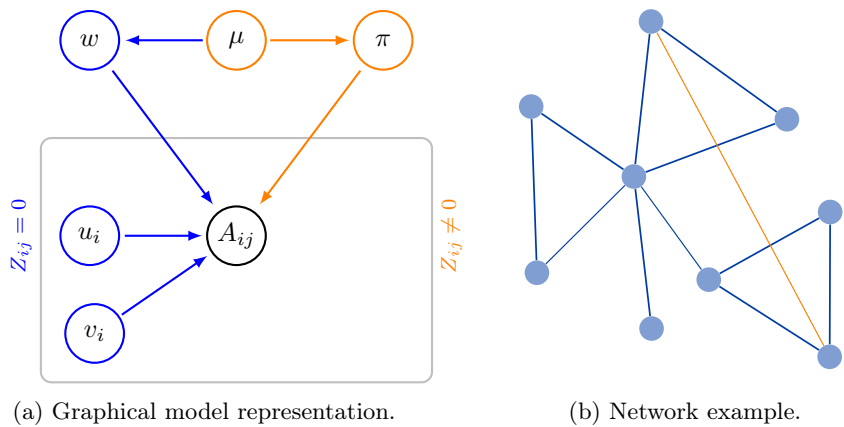

In this work, we aim to build our model upon recent developments in community detection \cite{de2017community} to address anomaly detection in networks. Specifically, we aim at incorporating  latent variables that measure the extent to which edges are classified as  anomalous, together with latent variables for the hidden community structure. More specifically,  by starting with an expressive generative model that captures the interaction patterns observed in network datasets, we can improve predictive power in detecting anomalous network interactions as well. The task is to infer both types of latent variables, i.e., communities and anomalies. 

We tackle the problem by building the core foundational probabilistic generative model, while  considering  the existence of individual anomalous edges. Our model outputs labels for edges, identifying them as legitimate or anomaly.  
We assume  that the only data we observe is the set of edges, coded by the adjacency matrix of entries $A_{ij}$, containing the weight of an edge between nodes $i$ and $j$.  Our goal is to determine which of the two possible mechanisms generated the edge and to label anomalies accordingly. Our approach is applicable to both directed and undirected networks. We present an efficient and scalable algorithm which could be easily  utilized by  practitioners  on networked datasets, without the need for extra node metadata. 

\section{Methods}  
\label{sec:LK} 
\subsection{Modeling anomalous  edges}
 
To achieve our goal of identifying anomalous edges and detecting communities in networks simultaneously, we need to explore statistical patterns in the connectivity
of the networks dictated by community structure. This can be obtained using the formalism of network probabilistic generative models \cite{goldenberg2010,ball2011efficient,de2017community}, as they are based on a rigorous theoretical framework and have efficient numerical implementations. These approaches assume that nodes are assigned to latent variables representing communities and these   community memberships determine the probability that edges exist between   the nodes. In particular, to model non-anomalous (or regular) edges, we consider the ideas proposed in \cite{de2017community}; as they flexibly apply to various types of networks with the characteristics needed in our problem: undirected and directed, weighted and unweighted, and it assumes mixed-membership community structure where nodes can belong to multiple communities. \\
We further assume that individual edges can be identified as anomalies when they deviate from what we consider a regular behavior, as described above. To model this, we introduce a binary random variable $Z_{ij} \in \ccup{0,1}$: when $Z_{ij}=1$  the edge $(i,j)$ is an anomaly. This is a latent variable that is not known in advanced and needs to be learned from data. It determines the probability distribution from where the edge $(i,j)$ is then extracted. 
From a generative modeling perspective, this setting can be understood as
first drawing latent labels on edges, $Z_{ij}$, that determine which edges are  anomalous and which edges are regular. Then drawing interactions $A_{ij}$ between nodes from a specific
distribution depending on the edge type, anomalous or regular.  A schematic representation  of our model is shown in Figure .1.   
Formally, the generative model is:
\begin{align}
Z_{ij} &\sim \bern(\mu) \label{eqn:Zprior}\\
A_{ij} &\sim \begin{cases}  \pois(A_{ij};\pi) & \text{if} \quad Z_{ij}=1 \quad \text{(\textit{anomalous edge})} \\ \pois(A_{ij};M_{ij}) & \text{if} \quad\ Z_{ij} = 0\quad \text{(\textit{regular edge})} \end{cases} \label{eqn:Poiss}\quad,
\end{align}
where $\mu \in \rup{0,1}$ is an hyper-parameter controlling the prior distribution of $Z$. Here, we assume a Poisson distribution with mean value of  $M_{ij}$ for the formation of regular edges, and a Poisson distribution with mean value of $\pi$ for the anomalous edge formation.  The parameter $M_{ij} = \sum_{k}u_{ik}v_{jk}w_{k}$ is controlled by community structure as $u_i,v_{i}$ are community membership vectors; $u_i = [u_{ik}]$ determines how much $i$ belongs to community $k$ considering the amount of \textit{out-going} edges; $v_i = [v_{ik}]$ only considers \textit{in-coming} edges. An affinity matrix $w$ of positive real-valued entries and dimension $K \times K$, where $K$ is the number of communities, encodes the density of edges in different communities, i.e., it shapes assortative or disassortative structures of the communities. Here we assume an assortative structure where nodes are more likely to exist within rather than between communities. This implies that $w=[w_{k}]$ is diagonal. However, similar derivations can be found for other types of structures. 
Collectively, we indicate with $\Theta=\bup{\ccup{u_{i}}_{i},\ccup{v_{i}}_{i},w,\pi, \ccup{Z_{ij}}_{i,j}}$ the latent variables of the model. 

\setlength{\textfloatsep}{6pt}
	\SetKwInOut{Input}{Input}
\begin{algorithm}[H]
 	\caption{\ACD \text{} : {EM algorithm.}}
	\label{alg:EM}
	\SetKwInOut{Input}{Input}
	\setstretch{0.7}
	\raggedright
	\Input{network $A=\{A_{ij}\}_{i,j=1}^{N}$, \\number of communities $K$.}
  	\BlankLine
	\KwOut{membership $u=\rup{u_{ik}},\, v=\rup{v_{ik}}$; network affinity matrix $w=\rup{w_{k}}$; mean value of Poisson anomaly distribution $\pi$;  prior on anomaly indicator $\mu$.}
	\BlankLine
	 Initialize $\Theta:(u,v,w,\pi, \mu)$ at random. 
	 \BlankLine
	 Repeat until $\mathcal{L}$ convergence:
	 \BlankLine
	\quad 1. Calculate $\rho$ and $Q$ (E-step): 
	\bea
	 &&\rho_{ijk}=u_{ik}v_{jk}w_{k}/\sum_{k}u_{ik}v_{jk}w_{k}  \;,\quad \nonumber \\ &&Q_{ij} = \f{\pois(A_{ij};\pi) \pois(A_{ji};\pi)\, \mu}{\pois(A_{ij};\pi) \pois(A_{ji};\pi)\, \mu+\pois(A_{ij};M_{ij})\,\pois(A_{ji};M_{ji})\, (1-\mu)}  \;.\quad  
	\nonumber
	\eea
	 \quad 2. Update parameters $\Theta$ (M-step):  
	\BlankLine
	\quad \quad \quad \quad 
		i) for each node $i$ and community $k$ update memberships:
		\bea
		\quad  u_{ik} = \f{a-1+\sum_{j} (1-Q_{ij})\, A_{ij}\rho_{ijk} }{b+\sum_{j}(1-Q_{ij})\, v_{jk}w_{k}} \nonumber\\
	    \quad  v_{jk} = \f{a-1+\sum_{i} (1-Q_{ij})\, A_{ij}\rho_{ijk} }{b+\sum_{i}(1-Q_{ij})\, u_{ik}w_{k}}  \nonumber
		\eea
	\quad \quad \quad \quad
	ii) for each community $k$ update affinity matrix:
		\be
		\quad w_{k} = \f{\sum_{i,j} (1-Q_{ij})\, A_{ij}\rho_{ijk} }{\sum_{i,j}(1-Q_{ij})\, u_{ik}v_{jk}} \quad. \nonumber
		 \ee
	\quad \quad \quad \quad
		iii) update anomaly parameter:
		\be \label{eqn:pi}
		\pi = \f{\sum_{i,j}Q_{ij}A_{ij}}{\sum_{i,j}Q_{ij}}.\nonumber
		\ee
		\quad \quad \quad
		iv) update  prior on anomaly indicator:
		\be \label{eqn:mu}
		\mu = \f{1}{N(N-1)/2}\sum_{i<j} Q_{ij} \quad.\nonumber
		\ee
		\quad \quad \quad

\end{algorithm}

Based on this, the probability of an edge $A_{ij}$ given the latent variables $\Theta$ can be written as:    
\be \label{eq:prob}
P(A_{ij}|\Theta) = \pois(A_{ij}; M_{ij})^{1-Z_{ij}} \,\pois(A_{ij}; \pi)^{Z_{ij}}\;.
\ee 
We assume a non-informative prior for $w$ and sparsity-enforcing priors for the membership vectors $u_{i},v_{i}$, thus encouraging the model to limit the number of non-zero entries.
 
Our goal is to estimate the latent variables, $\Theta$, given the adjacency matrix $A_{ij}$. To this end, we perform the inference task by maximizing the log-likelihood, $L( \Theta)=\log P(A|\Theta)$ with respect to $\Theta$. Given network data as the input, the desired output is inferring the probability that an edge is anomalous, as well as the underlying community structure, i.e., clustering nodes in communities. Our approach is capable of both learning how nodes are divided into groups and identifying those edges that are likely to be anomalous.  We implement  the inference task  using an Expectation-Maximization (EM) scheme as detailed in the Supplementary Material \Cref{appendix:edges_anom}. A pseudo-code of the algorithm is provided in \Cref{alg:EM}.  We refer to our model for anomaly detection in networks with community structure as \ACD.

The computational complexity of the algorithm scales as $O(E K + N^2)$, where $E$  is the total number of edges. In most of the applications, $K$ is usually much smaller than $E$ and  for sparse networks, as is often the case for real datasets,  $E \propto N$. Hence, the complexity is dominated by $O(N^2)$. This contribution comes from terms containing  $Q_{ij}$  that are not also multiplied by $A_{ij}$, i.e. terms in the denominators of the updates in \Cref{alg:EM}.  The matrix $Q$ is  a dense object that is necessary for classifying edges as anomalies. This may make it prohibitive to run our model on large systems. Exploring possible approximation to it to allow scaling to larger sizes is an interesting direction for future work.

\paragraph*{}
\section{Results on synthetic data}    
\label{sec:resSyn}
In order to validate the performance of our model and investigate its applicability, we apply it to synthetic datasets sampled with our generative model, see Supplementary Material for details (\Cref{appendix:GM}). These have known ground truth community memberships and anomalous edges. Hence, we assess the ability of  our algorithm to identify anomalous edges and in detecting communities. Once parameters are inferred, we use point estimates of $u_{i},v_{j}$ to assign nodes to group and of $Z_{ij}$ to classify edges. We compare these estimates with their respective ground truth values. As performance metrics we consider the F1 score  and cosine-similarity (CS), respectively. We are interested in particular in assessing how these quantities vary with $\rho_{a}$, the fraction of anomalous edges over the whole set of edges.
Specifically, we generate synthetic data sets with $N = 500$ nodes, average degree of $\langle k \rangle=20$, $K = 3$ hard communities of equal size with assortative structure and a range of $\rho_{a}\in \rup{0,1}$. 

As a baseline model for comparison, we consider a version of our model that reduces to standard community detection (CD) as in \cite{de2017community}. This is obtained by setting $\mu, \pi=0$ which are kept fixed as hyper-parameters in inference.
 
\begin{figure}[h!]  
  \caption{\large{\textbf{Community detection and anomaly detection in synthetic networks.} } a) Cosine similarity (CS) between ground truth and inferred communities and   
  b) the F1 score between the ground truth and the inferred anomalies. The model's ability to detect anomalous edges increases with increasing the ratio of anomalous edges in the network.  Synthetic networks with $N = 500$ nodes, $\pi=0.6$,  and $K = 3$ communities of equal-size unmixed group membership generated with \ACD \text{}. Lines and shades around them are averages and standard deviations over 20 sampled networks.}
	\includegraphics[width=0.9\linewidth]{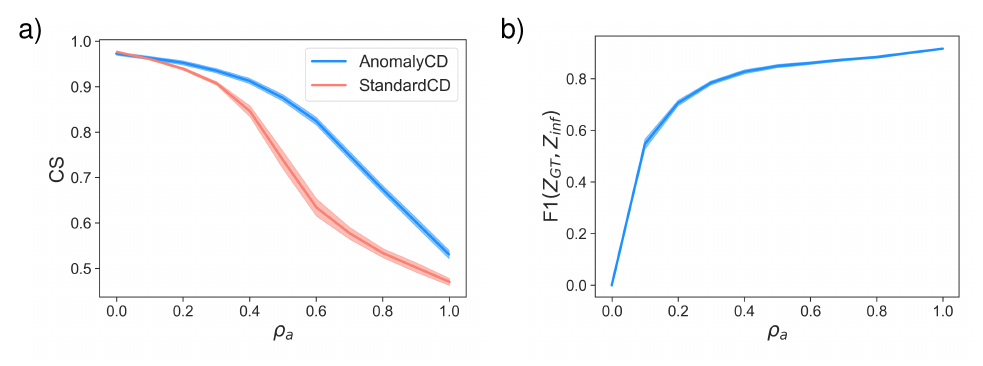}   
	\label{fig:CSsynth}
\end{figure}

We observe that  \ACD \textit{} significantly outperforms CD in detecting communities robustly across different values of $\rho_a$, as shown in  \Cref{fig:CSsynth}. In particular, its performance is stronger within an intermediate region where $\rho \in [0.4,0.6]$, i.e. when the majority of edges switches from being regular to being anomalous and CD's performance decays much faster.    In terms of anomaly detection, we observe that the performance improves by the increase of the anomaly density, with the largest improvement achieved for small values of $\rho_{a} < 0.2$, before reaching a steady increase towards the maximum value of 1 for larger $\rho_{a}$. 

\section{Results on Real World Datasets} 
\label{sec:RD}
In order to verify the validity of the algorithm and evaluate its performance on real-world datasets, we carry out three experiments. We study three real-world datasets with  node attributes available as potential  ground truth for comparison for community membership of nodes. More details on the studied datasets are available in  \Cref{appendix:data_desc}.  

\subsection{Experiment 1: injection of anomalous edges}

In the first experiment, we inject  anomalous edges in a given input network. These edges are selected uniformly at random among all the possible pairs of nodes that are not already connected by an edge. Then, we apply our method on this altered network and measure the algorithm's performance in i) detecting the injected edges, i.e., anomalous edges,  and ii)  detecting how communities are correlated with the node attributes available with the dataset. As performance metrics we measure precision, recall, and the Area Under the Curve (AUC) for anomaly detection and CS for community detection.  We vary the fraction $\rho_{a}$ of injected anomalous edges to assess how performance is impacted by this number.

\begin{figure}[h!] 
 \caption{\large{\textbf{The network of \polbooks \text{ } with injected edges (Experiment 1)}.} We compare the performance of the \ACD \text{} in terms of community detection and the detection of the anomalous edges injected in the network. We tune the percentage of injected edges $\rho_{a}$. Here $\pi=0.25$. Markers and errors denote means and standard deviations over  $20$ samples. }\label{fig:injection_polbooks}
	\includegraphics[width=0.9\linewidth]{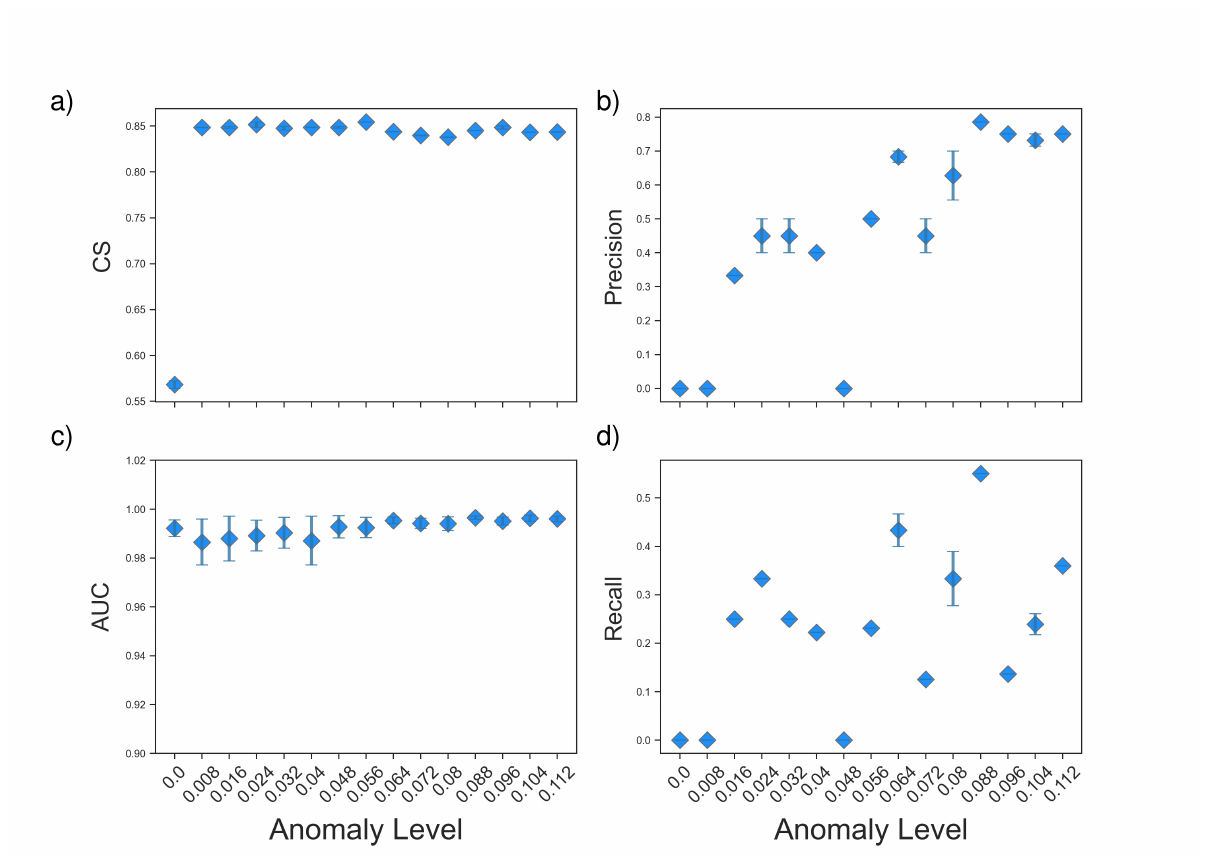}   
\end{figure}

\begin{figure}[t] 
	\includegraphics[width=0.9\linewidth]{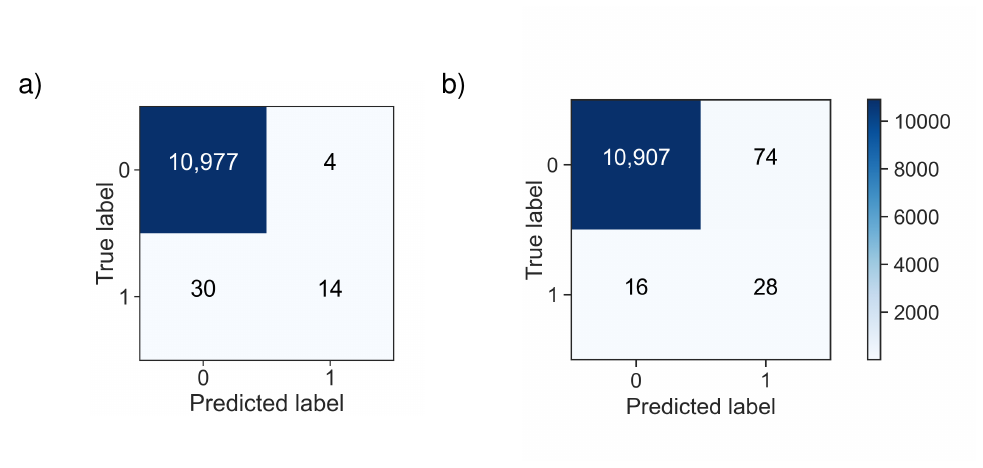}  
	\caption{\large{\textbf{The network of \polbooks \text{}  with injected edges (Experiment 1)}}. We compare the performance of \ACD \text{} in detecting the anomalous injected edges by estimating the confusion matrix for tow different values to initialize $\mu$. By adjusting the value of $\mu$, we can tune the recall and precision. a) $\mu=0.5, \pi=0.25$, b) $\mu=0.75, \pi=0.25$. }
	\label{fig:polbooks_CM}   
\end{figure}

\paragraph{Books about US politics}
The  network we study in this experiment contains $105$  books about  US politics which were published around the $2004$ presidential election \cite{KONECT} (\polbooks). In this network, nodes are books and an undirected, binary edge between two books indicates that those were co-purchased by the same costumer, for a total of $441$ edges. Injected anomalies here represent books that are either mistakenly co-purchased or mistakenly accounted in the dataset. \\
The results of this experiment  is presented in  \Cref{fig:injection_polbooks}. While AUC and community detection are both robust against  the number of edges injected in the network, precision and recall are more nuanced. This is due to the possibility of tuning the prior of $Z_{ij}$ via $\mu$  in order to obtain different regimes in retrieving anomalous edges. As can be  seen in  \Cref{tab:CM_polbooks} and \Cref{fig:polbooks_CM}, for a given level of injected anomalies, we can have high precision or high recall, depending on the initialized value of the prior. Hence, classification performance can be tuned towards either high precision or high recall by calibrating  $\mu$ accordingly, depending on the practitioner's goal. \\
For instance, when a practitioner wants to be strict in the criteria of labeling an edge as anomalous, thus avoiding labeling as ``anomalous'' edges that are not, then one should be more conservative with the prior, i.e. select a smaller $\mu$. Instead, when the priority is to detect as many anomalies as possible (at the risk of mislabeling true edges) one should increase $\mu$ and thus increase recall. This choice should depend on the application at hand, in particular one should reflect on the potential cost of classifying an edge as anomaly when it is not and compare it with the potential cost of missing anomalous edges.  \\

\begin{table}[t] 
\normalsize  
\caption{\large{\textbf{The confusion matrix for the network of \polbooks  \text{ } with injected edges (Experiment 1).}} We show how our model performs in terms of identifying anomalies--edges that have been injected in the dataset-- as we vary the prior on $Z$, tuned by the parameter $\mu$.  Here $\pi=0.25$ and $\rho_{a} = 0.087$. } 
\begin{tabular}{llcccc}
        \hline
 \textbf{ }  & $\mu$& \textbf{Precision}    & \textbf{Recall}  & \textbf{F1} \\  \hline
High Precision & $0.5$   & $0.78$      & $0.32$       & $0.45$   \\ 
High Recall& $0.75$  & $0.27$   & $0.64$  & $0.38$ \\  \hline 
\end{tabular}%
\label{tab:CM_polbooks}
\end{table}

\subsection{Experiment 2: 2-step inference of communities} 

In the second experiment we are interested in exploiting the information learned about anomalous edges to enhance performance in community detection. The hypothesis is that the presence of anomalies may corrupt community detection, for instance when anomalous edges connect two nodes that should not be part of the same communities. Using our model, we can act on the dataset by removing those edges that have higher probability of being anomalous, thus reducing noise in favor of better community detection. In practice, this is executed using a 2-step routine where we first run \ACD \textit{}  on the input dataset to estimate $Z_{ij}$. Then, we remove those edges with higher probability of being anomalous. Finally, we perform regular CD (running \ACD \textit{} with $\pi=\mu = 0$) on the ``cleaned'' network to extract communities.  We observe enhanced results in  the community detection task after removing the anomalous edges.

\paragraph{Zachary's Karate Club}  
We first test this second experiment on the dataset of  Zachary's Karate Club to convey in more qualitative terms how model improves upon the community detection task. The network's small size of $34$ nodes allows to better visualize the problem and how the 2-step routine works.
This social and undirected network shows the interactions between members of a university karate club for a period of three years from $1970$ to $1972$. The members are the nodes and an edge between a pair of members indicates  social interactions between them. During this period, due to administrative issues, the club splintered into two.  We exploits the membership of the nodes in the new clubs as possible meaningful  ground truth communities.\\
It is should be noted that what we refer to as the ground truth communities are in fact metadata of nodes that could be utilized to compare  the resulting communities.  The intention is merely to have a criterion for a quantitative comparison. In other words, we examine how the communities inferred by the algorithm are consistent with existing metadata.  In all the real data studied in this work, ‌ this interpretation of the metadata as the ground truth is applied.   \\ 
 
\begin{figure}[h!] 
 \caption{\large{\textbf{The network of Zachary karate Club (Experiment 2).}} a) The communities inferred by \ACD \text{}. The edges inferred by our algorithm as the anomalous edges are shown in red. Here, $\pi=10^{-4}, \mu=0.1$. b) The communities inferred by CD \text{}(i.e. \ACD \textit{} with $\pi=\mu = 0$).  c) Ground truth communities.}
	\includegraphics[width=1\linewidth]{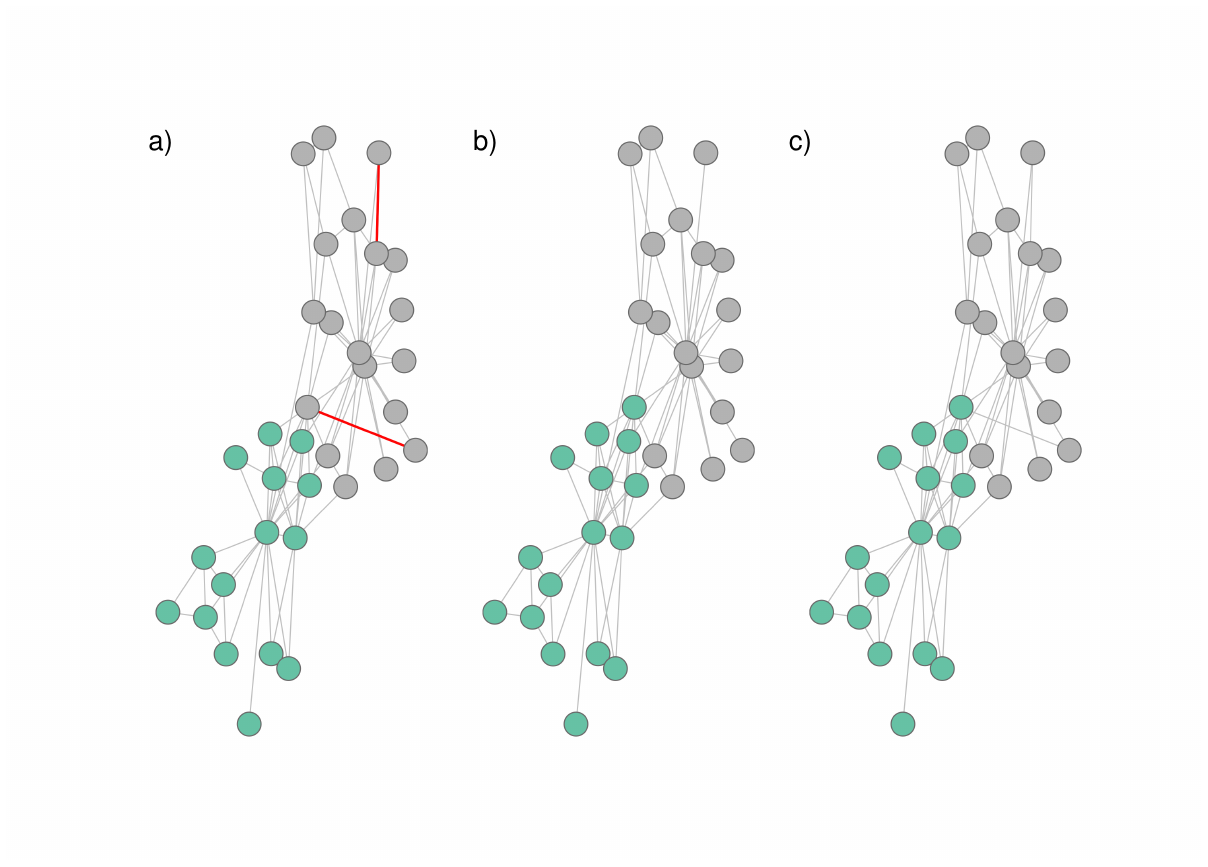}  
\label{fig:zachary_net}
\end{figure}

Figure \ref{fig:zachary_net} provides a visual representation of how the 2-step routine works. Communities inferred before and after removing anomalous edges are compared against those obtained using node attributes as ground truth.   We remove two edges classified with the highest probability as anomalous, these are shown in red in \Cref{fig:zachary_net}a. Removing the red edge connecting two more central nodes has the effect of changing the community assigned to one of the two nodes, which is now aligned with the ground truth after running CD the second time. Instead, the other node in the removed edge keep its community as detected in the first step, which was already aligned with the ground truth. As a result, the 2-step routine improves community detection performance. Notice also that removing the other anomalous edge does not impact performance, as the nodes involved in that edge do not switch communities and are already aligned with the ground truth. Hence, not all the removed edges may necessarily impact community detection the same.

\begin{figure}[h!] 
  \caption{\large{\textbf{The network of \polbooks \text{}  (Experiment 2).}} We show an example output of running the 2-step algorithm where we first detect anomalous edges, then remove them and finally run regular CD on the pruned dataset. ) The communities inferred by  \ACD \text{} before removal. The edges inferred by \ACD \text{} as anomalous edges are shown in red.  Here, $\pi=0.0001, \mu=0.5$. b) The communities inferred by CD after removal of anomalies. c) Ground truth  communities. Red rectangles  denote pairs of nodes that are connected by the edges inferred  as anomalous.  Cyan rectangles demonstrate the nodes that changed their community membership, after removing the aforementioned anomalous edges.  Cyan edges in panel (a) present the edges connected to these. }
	\includegraphics[width=1\linewidth]{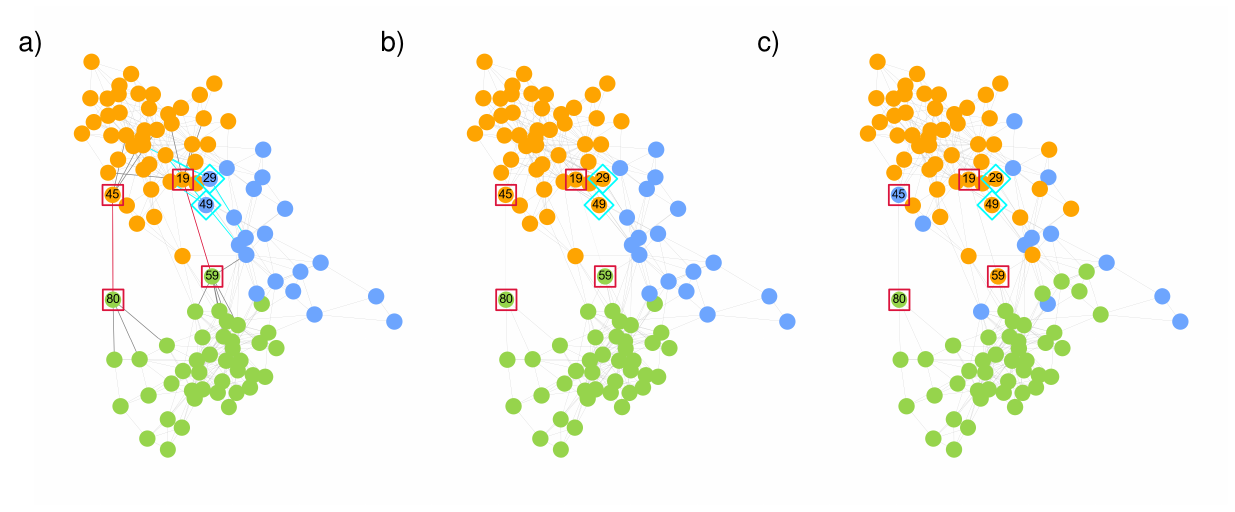}   
\label{fig:remove_polbooks} 
\end{figure}

We can now proceed analogously to apply the experiment on a larger dataset and present  quantitative results on the \polbooks \text{} dataset. 
Figure \ref{fig:remove_polbooks} demonstrates the communities inferred by  \ACD \text{} using the 2-step routine and the inferred anomalous edges are shown in red in \Cref{fig:remove_polbooks}a as in the example before. We notice that  the majority of the detected edges are between different communities. Comparing the communities detected in panels (a) and (b) with the ground truth communities in panel (c) we notice how the 2-step routine infers communities more aligned with the ground truth, as several nodes at the center of the figure switch communities from blue to green after removal.  In other words,  \ACD \textit{} is capable of improving community detection by uncovering edges that interfere with the community detection process.  \\     
 
Quantitatively, this is shown in \Cref{tab:CS_remove_polbooks} by an increase of CS value from 0.77 to 0.84 after anomaly removal, consistently over different values of prior's parameter $\mu$.   \Cref{tab:CS_remove_polbooks} emphasizes the robustness of the model in detecting communities  that are aligned with metadata information with respect to  changing the initial value of $\mu$.  In order to compare the performance of \ACD \textit{} in community detection task, we applied Bayesian Poisson matrix factorization (BPMF) \cite{gopalan2015scalable} on the \polbooks \text{} dataset, which results in $\text{CS} = 0.778\pm0.065$.  In addition,  in the table we report also the results of link prediction tests for model validation in the absence of ground truth (metadata are only a candidate for ground truth, true model parameters are unknown in real data). Specifically, we run 5-fold cross-validation and measure the AUC on the test dataset. Higher values indicate better performance in predicting missing edges, and better model's expressiveness. We see that the AUC only slightly drops when removing the anomalous edges, thus suggesting that removing information in an informed way (i.e. anomalous edges as detected by our model) enhances community detection without drastically affecting the AUC.   \\

\begin{table}[h!]  
\caption{\textbf{The network of \polbooks \text{} (Experiment 2)}. We present the ability of \ACD \text{} in  community detection, represented by cosine similarity (CS). Moreover, we validate the model by measuring the AUC in link prediction tasks. The results are robust with respect to the initial values of $\mu$. By removing the edges detected as anomalies, the  community detection task is improved.  Here, $\pi = 10^{-3}$.  The CS  errors are calculated by averaging over $20$ runs of the edge removal routine. To estimate AUC we perform 5-fold cross validation and report the averages and standard deviations over the 5 folds. } 
\begin{tabular}{cc|c|c|c|c} 
\cline{3-5}
& & \multicolumn{3}{ c| }{$\mu_0$} \\ \cline{1-5}
\multicolumn{1}{ |c| }{Measure} &Removed Anomalous Edges & $0.01$  &$0.05$   & $0.5$ \\ \cline{1-5}
\multicolumn{1}{ |c  }{\multirow{2}{*}{CS} } &
\multicolumn{1}{ |c| }{Yes} & $0.837 \pm 0.015$ & $0.834 \pm 0.020$  & $0.838 \pm 0.014$      \\ \cline{3-5}
\multicolumn{1}{ |c  }{}                        &
\multicolumn{1}{ |c| }{No} & $0.767 \pm 0.039$ & $0.766\pm 0.040$ & $0.768 \pm 0.040$       \\ \cline{1-5}
\multicolumn{1}{ |c  }{\multirow{2}{*}{AUC} } &
\multicolumn{1}{ |c| }{Yes} &$0.882	 \pm 0.018$ & $0.883 \pm 0.019$ & $0.888 \pm 0.020$   \\ \cline{3-5}
\multicolumn{1}{ |c  }{}                        &
\multicolumn{1}{ |c| }{No} &$0.953 \pm 0.058$ & $0.936 \pm 0.056$ & $0.922 \pm 0.045$    \\ \cline{1-5}
\end{tabular}
\label{tab:CS_remove_polbooks}   
\end{table}

\subsection{Experiment 3: Adding anomalous non-edges}

The first two sets of analyses examined the ability of the algorithm in detecting the anomalous edges and the impact of removing those edges from the dataset in community detection and link prediction tasks. However, our model is also able to estimate the probability of a non-edge to be anomalous. This can be used for instance in cases where we expect certain connections between nodes to happen, and if they are not observed we can use our algorithm to detect potential missing edges. Hence, we design a new experiment to  assess the possibility of improving the community detection task by  adding edges between disconnected nodes. In other words,  the algorithm detects unseen edges  which may improve community detection if we were to add them to the network, in a similar 2-step routines as before, this time by adding instead of removing edges.

As in the previous experiments, we apply \ACD \textit{} on the dataset to estimate $Z$. However, in this case, we select the entries corresponding to non-edges (i.e. $A_{ij}=0$) which have highest probability of being anomalous and then add them to the network. Then, we apply regular CD on the final dataset and compare communities inferred  before and after adding these edges.

\paragraph{American college football}
The experiment was tested on a network of football games between American colleges in the fall of $2000$ \cite{Girvan2002}. Nodes in the network indicate the college teams and the undirected, and binary edges connecting them represent the games between the teams. 
The teams are divided into $12$ conferences where members of each conference have more frequent games with each others compared to the games with members of other  conferences. We use the conference membership as candidate ground truth community memberships to compare against. An anomalous non-edge corresponds to a game that has not been planned by the league's organizers but could have been played  (for instance by adding more games to the fixtures or substituting with other games currently in the fixture), as it aligns with the pattern of existing games.  In this context, selecting the fixtures is an important task for the organizers, as the set of matches that a team has to play can significantly impact its chance to go to the final National Championship. \\
  
\begin{figure}[h!] 
 \caption{\large{\textbf{The network of American college football  after addition of edges (Experiment 3).}} a) The communities inferred by CD   on the original input data.  b) Community detection by CD algorithm on the dataset after addition of the potential inferred  edges  inferred   by \ACD \text{} (these edges are shown in red). c) Ground truth communities. The nodes that changed their communities between a) and b) are shown with cyan border. Those with red border are nodes connecting to added edges. The name of some of the nodes impacted by the edge addition is added to the figure as the footnote.  }
	\includegraphics[width=1.1\linewidth]{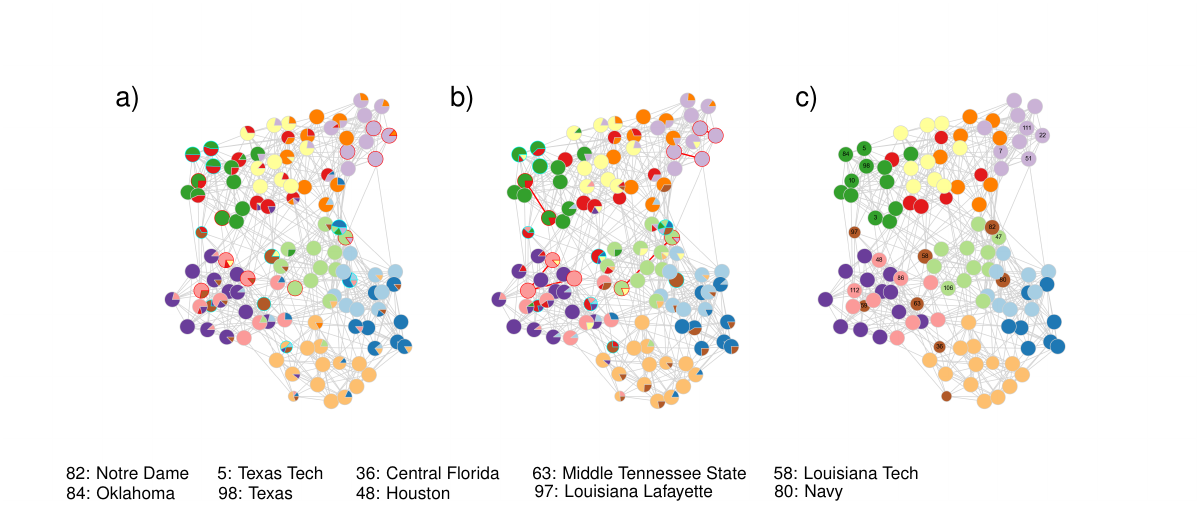}   
\label{fig:football_net} 
\end{figure}

In \Cref{fig:football_net} we show a qualitative example of how communities change before and after  addition of 6 non-existing edges classified with the highest probability of being anomalous ( they correspond to $1\%$ of the total number of edges). It is clear from this figure that addition of few edges impacts the community membership of several nodes, and not only those directly connected to the newly added edges. In particular, it strengthen the memberships of nodes in the communities of the nodes directly connected to the added edge (see e.g. the green and pink, whose nodes become less overlapping). In addition, it softens the membership of several nodes that are ``Independent'' (brown nodes in \Cref{fig:football_net}c), they do not belong to any conference. These are nodes that play several games against nodes in various conferences. 
 In general both approaches achieve strong results in detecting communities that align with conference membership. However, the 2-step routine with edge addition improves performance further, as detected by both CS and F1-score, see \Cref{tab:hidden_football}.  These results show the flexibility of our model in detecting various types of anomalies and acting on them by suitably modifying the network to enhance community detection.

\begin{table}[h!]
\caption{\large{\textbf{The network of American college football (Experiment 3)}}.  We report CS and F1-score between the inferred communities and those given by metadata. Averages and standard deviations are over $20$ runs of the edge addition routine.}
      \begin{tabular}{cccc} 
        \hline
           & CD on dataset  &\ACD \text{} on dataset   & CD on dataset with added edges\\ \hline
        CS &0.955 $\pm$ 0.003& 0.957 $\pm$ 0.001 & 0.957 $\pm$ 0.003\\
        F1 & 0.959 $\pm$ 0.005& 0.957 $\pm$ 0.003  & 0.962 $\pm$ 0.006\\ \hline
      \end{tabular} 
\label{tab:hidden_football}
\end{table}

\section{Discussion}   
We have proposed a probabilistic model for detecting anomalies in networks. It relies on the assumption that regular patterns of interactions are determined by community structure and exploits this insight to detect pairs of nodes, existing or non-existing ties, that deviate from regular behavior.\\ 
The algorithmic implementation uses an expectation-maximization routine that outputs both community membership of nodes and probability estimates for pairs of nodes of being anomalous or not. We find that in synthetic data it improves community detection while also showing robust performance in identifying anomalies.\\ 
In addition, in the case of real-world datasets, we have performed various experiments that show an increase in the model's ability in community detection tasks. Specifically, both in the experiment where the inferred  anomalous edges were removed from the network, and in the case where non-existing but potential ties were identified and added to the network, there was an improvement in detecting communities that are aligned with metadata. Also,  in another experiment, in which anomalous edges were injected into the system, ‌ our model showed high capability in detecting these ties.\\      
We have focused here in anomalies on pairs of nodes, edges or non-edges but similar ideas and methods can be used to extend this model to anomalies on nodes. In this context, it may be interesting to explore future extensions that incorporate extra information, e.g.  node attributes, along with community structure, as done for instance in \cite{bojchevski2018bayesian,contisciani2020,newman2016structure,hric2016network}.
As accurately identifying anomalies is deeply connected with the chosen null model determining what regular patterns are, it is important to consider other possible mechanism for tie formation, beyond community structure.  
In recent works \cite{safdari2020generative,safdari2022reciprocity,contisciani2021community}, we found that modeling community patterns together with reciprocity effects, leads to higher predictive performance, thus more expressive generative models. This could significantly change the performance of our foundational model as well. Hence, a natural next step is to include reciprocity in our model and measure  how, by varying the intensity of these effects, anomaly detection improves (or decreases). \\
As a final remark, it is worth mentioning that what is referred to as an anomalous edge in this work should not necessarily be interpreted  as an undesirable interaction or malicious activity. Indeed, an anomaly here reflects an unusual pattern, as compared to that of other edges, which can not be explained by the core structural pattern of the dataset, in this case driven by community structure. 
We encourage practitioners to carefully assess its qualitative interpretation based on the application at hand and preferably guided by domain expertise and knowledge.

\section*{ Availability of data and materials}    
The data analyzed in this study are available at \cite{Girvan2002,KONECT,Zac77}. The code to run the model  is publicly available  at \url{https://github.com/hds-safdari/Anomaly_Community_Detection}.

\section*{Acknowledgements}
All the authors were supported by the Cyber Valley Research Fund.

\bibliographystyle{apsrev4-1} 
\bibliography{manuscript}
 
\clearpage
\newpage 
\section*{Supporting Information (SI)}
\subsection{Appendix 1: Inference with expectation-maximization} 
\label{appendix:inference}  
Our goal is, given  two mechanisms responsible for edge formation, first to determine the values of the  parameters $\Theta:(u_{ik},v_{ik},w_k,\pi),\mu$, which  determine  the  relationship between the anomaly indicator $Z_{ij}$ and the data, and then, given those values, to estimate the indicator $Z_{ij}$ itself.
We have the posterior:
\be\label{eqn:joint}
P(Z,\Theta| A) = \f{P(A|Z,\Theta) P(Z|\mu) P(\Theta)P(\mu)}{P(A)} \quad.
\ee
Summing over all the possible indicators we have:
\be
P(\Theta| A) = \sum_{Z}P(Z,\Theta| A) \quad,
\ee 
which is the quantity that we need to maximize to extract the optimal $\Theta$.
It is more convenient to maximize its logarithm, log-likelihood, as the two maxima coincide. We use the Jensen's inequality:
\be\label{eqn:jensen} 
L(\Theta) = \log P(\Theta| A) = \log\sum_{Z}P(Z,\Theta| A) \geq  \sum_{Z} q(Z)\, \log \f{P(Z,\Theta| A)}{q(Z)} \quad,   
\ee

where $q(Z)$ is a variational distribution that must sum to $1$. In fact, the exact equality happens when,
\be\label{eqn:q}
q(Z) = \f{P(Z,\Theta| A)}{\sum_{Z } P(Z,\Theta| A)}\quad,
\ee
this definition is also maximizing the right-hand-side of Eq.~(\ref{eqn:jensen}) w.r.t. $q$.\\
Finally, we  need to maximize the log-likelihood with respect to $\Theta$ to get the latent variables. This can be done in an iterative way using Expectation-Maximization algorithm (EM), alternating between maximizing w.r.t. $q$ using Eq.~(\ref{eqn:q}) and then maximzing Eq.~(\ref{eq:Lconv}) w.r.t. $\Theta$.

We start by derivation of Eq.~(\ref{eqn:jensen})  with respect to the individual parameters, for example we start  by considering $u_{ik}$.
We assume uniform prior w.r.t. $\Theta$, but we can easily incorporate more complex choices if needed.
\bea
\sum_{Z}q(Z) \f{\partial}{\partial u_{ik}} \rup{\log \f{P(Z,\Theta| A)}{q(Z)} }&=& \sum_{Z}q(Z) \f{\partial}{\partial u_{ik}} \log P(Z,\Theta| A) \\
&=& \sum_{Z}q(Z) \f{\partial}{\partial u_{ik}} \sum_{i,j}(1-Z_{ij}) \,\log \pois( A_{ij};M_{ij}) \\
&=& \sum_{Z,j}q(Z)\, (1-Z_{ij}) \, \f{\partial}{\partial u_{ik}}  \rup{-u_{ik}v_{jk}w_{k} +A_{ij}\log\sum_{k} u_{ik}v_{jk}w_{k}}\qquad \\
&=& \sum_{Z,j}q(Z)\, (1-Z_{ij}) \,  \rup{-v_{jk}w_{k} +A_{ij}\f{\rho_{ijk}}{u_{ik}}} =0 \quad,
\eea
where in the last equation we used once again Jensen's inequality with:
\be \label{eqn:rho}
\rho_{ijk}=u_{ik}v_{jk}w_{k}/\sum_{k}u_{ik}v_{jk}w_{k} \quad.
\ee
Defining $Q_{ij} = \sum_{Z} q(Z)\, Z_{ij}$, the expected value of $Z_{ij}$ over the variational distribution, we finally obtain:
\be \label{eqn:u}
u_{ik} = \f{\sum_{j} (1-Q_{ij})\, A_{ij}\rho_{ijk} }{\sum_{j}(1-Q_{ij})\, v_{jk}w_{k}} \quad.
\ee
We find similar expression for $v_{ik}$ and $w_{k}$,
\be\label{eqn:v}
v_{jk} = \f{\sum_{i} (1-Q_{ij})\, A_{ij}\rho_{ijk} }{\sum_{i}(1-Q_{ij})\, u_{ik}w_{k}} \quad,
\ee

\be\label{eqn:w}
w_{k} = \f{\sum_{i,j} (1-Q_{ij})\, A_{ij}\rho_{ijk} }{\sum_{i,j}(1-Q_{ij})\, u_{ik}v_{jk}} \quad.
\ee
For $\pi$:
\bea
\sum_{Z}q(Z) \f{\partial}{\partial \pi} \rup{\log \f{P(Z,\Theta| A)}{q(Z)} }&=& \sum_{Z}q(Z) \f{\partial}{\partial \pi} \log P(Z,\Theta| A) \\
&=& \sum_{Z}q(Z) \f{\partial}{\partial \pi} \sum_{i,j}\rup{Z_{ij}\, \log \pois( A_{ij};\pi)} \\
&=& \sum_{Z}q(Z)\,\f{\partial}{\partial \pi} \sum_{i,j}\rup{Z_{ij}\, (-\pi+A_{ij} \log \pi)}\,\quad \\
&=& \sum_{Z,i,j}q(Z)\,\rup{Z_{ij}\, (-1+A_{ij}\, \f{1}{\pi})}\, = 0\quad,
\eea
yielding
\be \label{eqn:pi}
\pi = \f{\sum_{i,j}Q_{ij}A_{ij}}{\sum_{i,j}Q_{ij}}.
\ee

Similarly for $\mu$:
\bea
\sum_{Z}q(Z) \f{\partial}{\partial \mu} \rup{\log \f{P(Z,\Theta| A)}{q(Z)} }&=&\sum_{Z}q(Z) \sum_{i<j} \f{\partial}{\partial \mu}  \rup{Z_{ij}\log\mu +(1-Z_{ij})\log(1-\mu)}\\
&=&\f{1}{ \mu} \,\sum_{i<j} Q_{ij} -\f{1}{1-\mu}\, \sum_{i<j} (1-Q_{ij}) \quad,
\eea
yielding:
\be \label{eqn:mu}
\mu = \f{1}{N(N-1)/2}\sum_{i<j} Q_{ij} \quad.
\ee

To evaluate $q(Z)$, we substitute the estimated parameters inside Eq.~(\ref{eqn:q}):
\bea
q(Z) &=& \f{\prod_{i,j}  \pois(A_{ij};\pi)^{Z_{ij}} \, \pois(A_{ij}; M_{ij})^{1-Z_{ij}} \,\prod_{i<j} \mu^{Z_{ij}}\, (1-\mu)^{(1-Z_{ij})}}
{\sum_{Z}\prod_{i,j}  \pois(A_{ij};\pi)^{Z_{ij}}\, \pois(A_{ij}; M_{ij})^{1-Z_{ij}} \,\prod_{i<j} \mu^{Z_{ij}}\, (1-\mu)^{(1-Z_{ij})}} \\
&=& \prod_{i<j} \f{ \rup{\pois(A_{ij};\pi) \pois(A_{ji};\pi)\, \mu}^{Z_{ij}}\,\rup{\pois(A_{ij};M_{ij})\,\pois(A_{ji};M_{ji})\, (1-\mu)}^{(1-Z_{ij})}}
{\sum_{Z_{ij}=0,1}  \rup{\pois(A_{ij};\pi) \pois(A_{ji};\pi)\, \mu}^{Z_{ij}}\,\rup{\pois(A_{ij};M_{ij})\,\pois(A_{ji};M_{ji})\, (1-\mu)}^{(1-Z_{ij})}}\qquad\\
&=& \prod_{i<j}\, Q_{ij}^{Z_{ij}}\, (1-Q_{ij})^{(1-Z_{ij})} \quad, \label{eqn:qQij}
\eea
where
\be\label{appendix:eqn:Qij}
Q_{ij} = \f{\pois(A_{ij};\pi) \pois(A_{ji};\pi)\, \mu}{\pois(A_{ij};\pi) \pois(A_{ji};\pi)\, \mu+\pois(A_{ij};M_{ij})\,\pois(A_{ji};M_{ji})\, (1-\mu)} \quad.
\ee
\subsubsection{Covergence criteria}
\label{appendix:edges_anom}
The EM algorithm consists of randomly initializing  $\pi, \mu, u,v,w$, then iteration of Eqs.~\ref{eqn:rho}-\ref{eqn:w},  \ref{eqn:pi}, \ref{eqn:mu}, and \ref{eqn:qQij}, until the convergence of the following log-posterior, 
\bea
L(\Theta) &=& \log P(\Theta|A) \geq {\sum_{Z} q(Z) \log \f{P(Z,\Theta|A)}{q(Z)}} \\
&=& -{\sum_{Z} q(Z) \log q(Z)}+{\sum_{Z} q(Z) \left\{ \log P(A|Z;\Theta)+ \log P(Z| \mu) + \log P(\Theta)+\log P(\mu)  \right\}} \\
&=&  -{\sum_{Z} q(Z) \log q(Z)}+ \log P(\Theta)+\log P(\mu) \\
&+& {\sum_{Z} q(Z) \left\{  {\sum_{i,j} \rup{Z_{i,j}  \log  \pois(A_{i,j}; \pi)+(1-Z_{i,j}) \log  \pois(A_{i,j}; M_{i,j})+Z_{i,j} \log \mu +(1-Z_{i,j})\log (1-\mu)} }  \right\}} \nonumber  \\
&=&  -\sum_{i<j} \rup{Q_{i,j}\log Q_{i,j} +(1-Q_{i,j}) \log (1-Q_{i,j}) } + \sum_{i,j}  \left\{ Q_{i,j}  \left( -\pi +A_{i,j}  \log \pi  \right)   \right.  \nonumber \\
&& \left. + (1-Q_{i,j} ) \left( -M_{i,j} +A_{i,j}  \log M_{i,j}  \right) +Q_{i,j}  \log \mu + (1-Q_{i,j} ) \log (1-\mu) \right\} +const \qquad,
\eea

where we neglect $const$, constant term due to the uniform priors. To calculate $\log q(Z)$, we used Eq.(\ref{eqn:qQij}), i.e., a Bernoulli distribution.

One can further add parameters' regularization, for instance by assuming Gamma-distributed priors for the membership vectors,
\be
P(u_{ik}; a,b) \propto u_{ik}^{a-1}e^{-b u_{ik}} \quad,  
\ee

where $a\geq 1$, to  ensure the maximization of the log-likelihood (the second derivative must be negative), similarly for the $v_{ik}$. This would add new terms to the log-likelihood:
\bea
\mathcal{L}(\Theta) &=& L(\Theta) + (a-1) \sum_{i,k}\log u_{ik} -b \sum_{ik}u_{ik} \nonumber\\
&&+ (a-1) \sum_{i,k}\log v_{ik} -b \sum_{ik}v_{ik}\quad. \label{eq:Lconv}   
\eea  
Alternatively, one can add constraints to the parameters, e.g. $\sum_{k}u_{ik}=1$ (and similarly for $v_{i}$). This would modify the likelihood by adding the corresponding Lagrange multipliers.

\section{Appendix 2: Generative model}  
\label{appendix:GM}  
Being generative, our model can be used to generate synthetic networks that include both  anomalous edges and community structure. To this end, we sample the parameters $(u,v,w,\mu,\pi)$ and then, given these latent variables, we sample $Z$. Finally,  given the $Z$ and the latent variables, we can sample the adjacency matrix $A$.

For a given set of community parameters as the input \cite{de2017community,contisciani2020}, we sample anomalous edges from a Poisson distribution as in \Cref{eqn:Poiss}, with a  Bernoulli prior  as in \Cref{eqn:Zprior}. The mean value of the Poisson distribution, $\pi$,  is constant for all edges, however, its value can be chosen in order to control the ratio $\rho_{a}$ of edges being anomalous over the total number of edges. The average number of anomalous  and non-anomalous edges are $N^2\mu\, (1-e^{-\pi})$, and $(1-\mu)\, \sum_{i,j}(1-e^{-M_{ij}})$, respectively. Assuming a desired total number of edges $E$, we can  multiply $\pi, \mu$ and $M_{ij}$ by suitable sparsity constants that tune: i) the ratio $$\rho_{a}=\f{N^2\mu\, (1-e^{-\pi})}{N^2\mu\, (1-e^{-\pi})+(1-\mu)\, \sum_{i,j}(1-e^{-M_{ij}})}, \quad \in \rup{0,1};$$ ii) the success rate of anomalous edges $\pi$. Once these two quantities  are fixed, the remaining sparsity parameter for the matrix $M$, is estimated as:
\be
E\,(1-\rho_{a}) = (1-\mu)\, \sum_{i,j}(1-e^{-cM_{ij}})\quad, 
\ee
which can be solved with root-finding methods.

\section{Appendix 3: Performance in real-world datasets}  
\label{appendix:RD} 

\subsection*{Real data: dataset description}
\label{appendix:data_desc} 
We tested our algorithm on three real-world datasets with the available ground truth communities. 
A brief overview of  features of the studied datasets is presented  in the \Cref{tabSI:data_desc}. 

\begin{table}[h!]
\caption{{\bf {Datasets description.}}} 
\begin{tabular}{llllll}
\hline
 \textbf{Network} & \textbf{Abbreviation}    & \textbf{N} & \textbf{E} & \textbf{Ref.}\\ \hline
American college football   \hspace{30pt}  & college football    \hspace{10pt}  & $115$   \hspace{10pt}     & $613$     \hspace{10pt}      & \cite{Girvan2002}    \\  
Political blogs network &POLBLOGS      & $1,490$   & $19,090$  &    \cite{Adamic_polblogs2005}    \\ 
Books about US politics & POLBOOKS      & $105$   & $441$  &    \cite{KONECT}    \\ 
Zachary karate club & Zachary     & $34$   & $78$ &    \cite{Zac77}    \\ \hline
\end{tabular}%
\label{tabSI:data_desc} 
\end{table}

\subsection*{Real data: performance} 
\begin{figure}[h!]
 \caption{\large{\textbf{The network of Zachary karate Club  with injected edges  (Experiment 1).}} Precision= $1.0$, recall= $0.667$, and F1 score= $0.8$. Here $\pi=0.25, \mu_0=0.5$.}
	\includegraphics[width=0.45\linewidth]{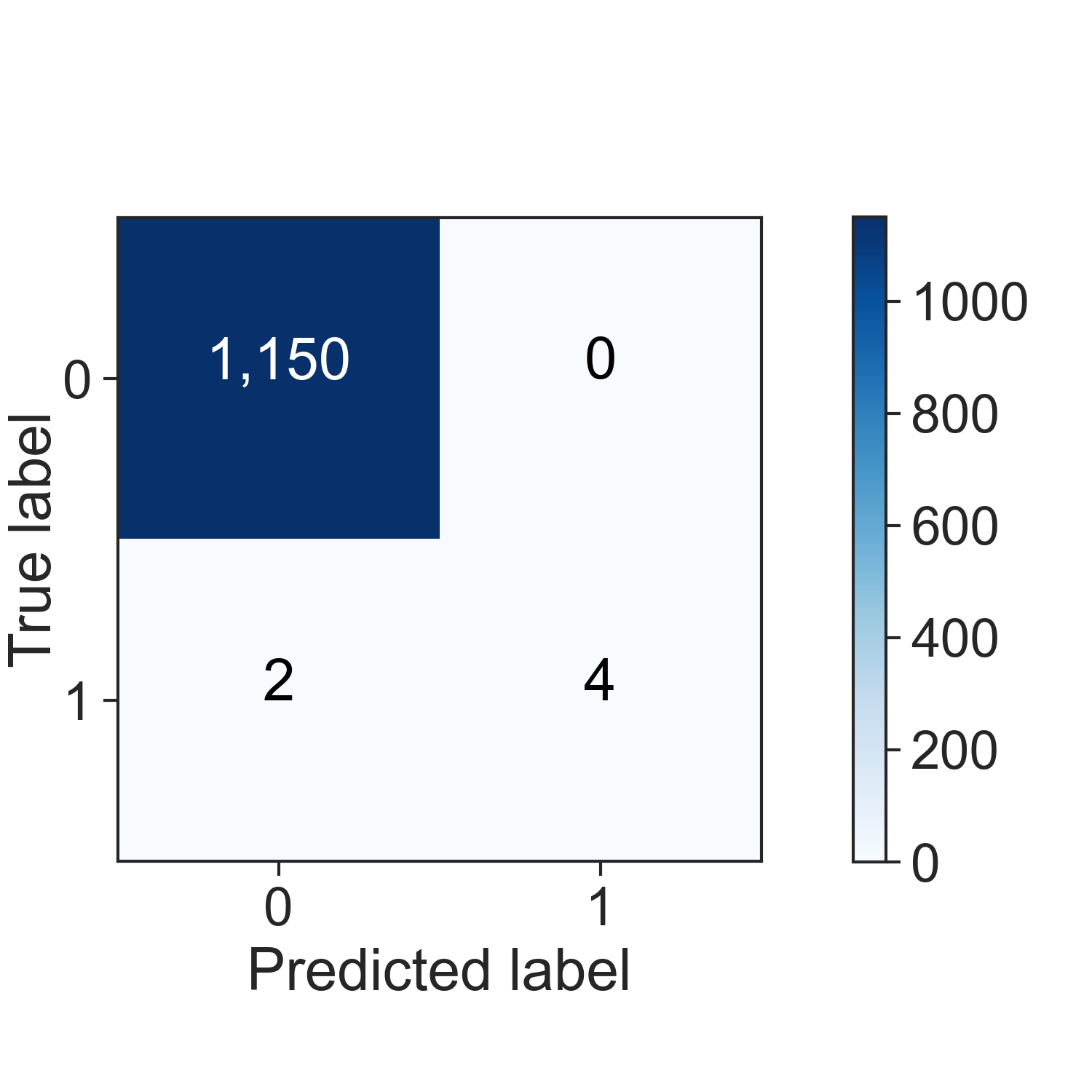}   
	\label{fig:CM_injection_zachary}
\end{figure}

\begin{figure}[htb]   
\caption{\large{\textbf{The network of Zachary karate Club (Experiment 1).}} a) $Q$ matrix, estimated by \ACD \textit{}, as the expected value of $Z_{ij}$ over the variational distribution. b) The inferred anomalous edges, i.e., the entries of $Q$ with the values above the threshold, here, the threshold = $0.7\times max(Q)$.  Here, $\mu = 0.1,
\pi = 10^{-4}$. } 
	\includegraphics[width=1.\linewidth]{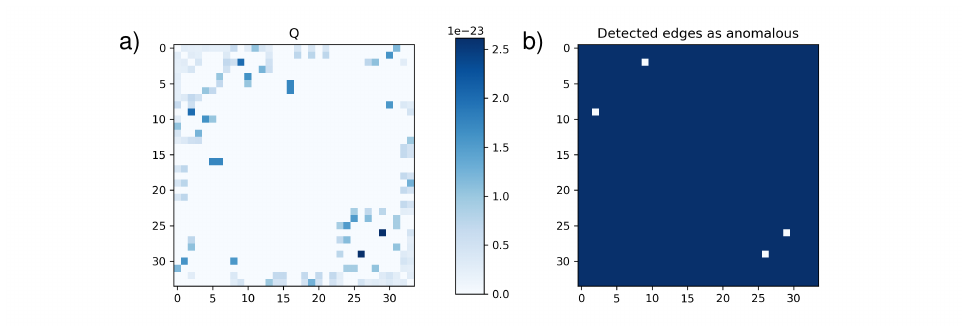} 
	\label{figSI:Q_zachary}
\end{figure}

Figure \ref{fig:CM_injection_zachary} shows how \ACD \textit{} can capture the  anomalous edges with a satisfying accuracy, by tuning the parameters of our model, i.e., $\mu$ and $\pi$. \\ 
We apply \ACD \textit{} on a network with injected anomaly to estimate $Q_{ij}$, as the  the expected value of $Z_{ij}$ over the variational distribution (see \Cref{appendix:inference}). The entries with  the highest values  are detected as anomalous edges, \Cref{figSI:Q_zachary}.

\begin{figure}[htb]    
\caption{\large{\textbf{The network of \polbooks \text{} ( Experiment 2).} }a) Soft community membership of nodes inferred by  \ACD \text{}. The edges inferred by \ACD \text{} as anomalous edges are shown in red.  Here, $\pi=0.0001, \mu=0.5$. b) The communities inferred by CD, after removing the edges  inferred as anomalous. c) Ground truth  communities.  Red rectangles  denote pairs of nodes that are connected by the edges inferred  as anomalous.  Cyan rectangles demonstrate the nodes that changed their community membership, after removing the aforementioned anomalous edges. Cyan edges in panel (a) present the edges connected to these.  } 
	\includegraphics[width=1.\linewidth]{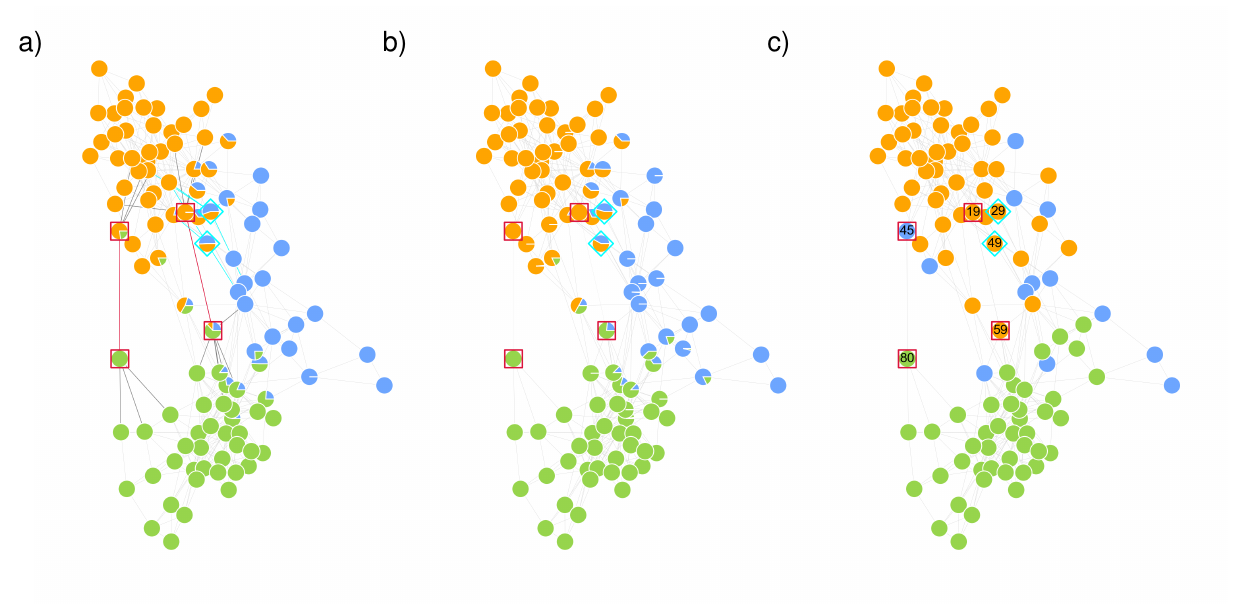} 
	\label{figSI:Polbooks_SCD}
\end{figure}

%
%

\paragraph*{} Political blogs:To evaluate the efficiency and ability of our model when appiled to  larger datasets, we used our model on the network of hyperlinks between weblogs on US politics \cite{Adamic_polblogs2005} . This is a directed network with $1490$ nodes, each belonging to one of  two categories: conservative, or liberal. We run the Experiment 1 on this dataset by injecting some random anomalous edges, then applying the algorithm to detect those injected edges. Depending on the aim of the study, we can tune  the value of the priors to have a higher precision or recall.  \Cref{tab:CM_polblogs}  displays different regimes of performance with respect to the value of anomaly parameter, $pi$.

\begin{table}[htb] 
\normalsize  
\caption{ \large{\bf{The confusion matrix for the network of  \polblogs \text{} with injected edges (Experiment 1)}}. We show how our model performs in terms of identifying anomalies--edges that have been injected in the dataset-- as we vary the anomaly parameters, $\pi$.  Here $\mu=0.5$ and $\rho_{a} = 0.33$. \\ } 
\begin{tabular}{llcccc}
        \hline
 \textbf{ }  & $\pi$& \textbf{Precision}    & \textbf{Recall}  & \textbf{F1} \\  \hline
High Recall/Precision & $ 0.3$   & $0.49$      & $0.99$       & $0.65$   \\ 
High Precision & $ 0.5$   & $0.45$      & $0.27$       & $0.34$   \\ 
High Recall& $0.8$  & $0.40$   & $0.69$  & $0.51$ \\  \hline 
\end{tabular}%
\label{tab:CM_polblogs}
\end{table}
 
\end{document}